\documentclass[a4paper,fleqn,usenatbib]{mnras}

\usepackage{newtxtext,newtxmath}

\usepackage[T1]{fontenc}
\usepackage{ae,aecompl}

\usepackage[usenames, dvipsnames]{color}

\usepackage{graphicx}	
\usepackage{amsmath}	
\usepackage{amssymb}	

\title[Variable Diffuse Interstellar Bands]{TRES Survey of Variable Diffuse Interstellar Bands}

\author[C.\ Law et al.]{Charles~J.~Law,$^{1}$\thanks{E-mail: claw@college.harvard.edu}
Dan~Milisavljevic,$^{1}$
Kyle~N.~Crabtree,$^{2}$
Sommer~L.~Johansen,$^{2}$\newauthor
Daniel~J.~Patnaude,$^{1}$
Raffaella~Margutti,$^{3}$
Jerod~T.~Parrent,$^{1}$
Maria~R.~Drout,$^{4, 5}$\newauthor
Nathan E.\ Sanders,$^{1}$
Robert P.\ Kirshner,$^{1, 6}$
David~W.~Latham$^{1}$
\\
$^{1}$Harvard-Smithsonian Center for Astrophysics, 60 Garden St., Cambridge, MA 02138, USA\\
$^{2}$Department of Chemistry, University of California, Davis, One Shields Ave., Davis, CA 95616, USA\\
$^{3}$Center for Interdisciplinary Exploration and
  Research in Astrophysics (CIERA) and Department of Physics and
  Astrophysics, \\ Northwestern University, Evanston, IL 60208, USA\\
$^{4}$The Observatories of the Carnegie Institution for Science, 813 Santa Barbara St., Pasadena, CA 91101, USA\\
$^{5}$Hubble, Carnegie-Dunlap Fellow\\
$^{6}$Gordon and Betty Moore Foundation, 1661 Page Mill Road, Palo Alto, CA 94304, USA\\
}

\date{Accepted XXX. Received YYY; in original form ZZZ}

\pubyear{2016}

\hypersetup{draft}
\begin{document}
\label{firstpage}
\pagerange{\pageref{firstpage}--\pageref{lastpage}}
\maketitle

\begin{abstract} 

  Diffuse interstellar bands (DIBs) are absorption features commonly
  observed in optical/near-infrared spectra of stars and thought to be
  associated with polyatomic molecules that comprise a significant
  reservoir of organic material in the universe. However, the central
  wavelengths of almost all DIBs do not correspond with electronic
  transitions of known atomic or molecular species and the specific
  physical nature of their carriers remains inconclusive despite
  decades of observational, theoretical, and experimental research. It
  is well established that DIB carriers are located in the
  interstellar medium, but the recent discovery of time-varying DIBs
  in the spectra of the extragalactic supernova SN\,2012ap suggests
  that some may be created in massive star environments. Here we
  report evidence of short time-scale ($\sim$10--60~d) changes in DIB
  absorption line substructure toward 3 of 17 massive stars observed
  as part of a pathfinder survey of variable DIBs conducted with the
  1.5-m Tillinghast telescope and Tillinghast Reflector Echelle
  Spectrograph (TRES) at Fred L. Whipple Observatory. The detections
  are made in high-resolution optical spectra ($R \sim 44000$) having
  signal-to-noise ratios of 5--15 around the 5797 and 6614~\AA\
  features, and are considered significant but requiring further
  investigation.  We find that these changes are potentially
  consistent with interactions between stellar winds and DIB carriers
  in close proximity. Our findings motivate a larger survey to further
  characterize these variations and may establish a powerful new
  method for probing the poorly understood physical characteristics of
  DIB carriers.

\end{abstract}




\section{Introduction}

The nature of diffuse interstellar bands (DIBs) remains a
long-standing problem in optical and near-infrared astronomy. DIBs,
which represent over 500 absorption features, are generally narrow
[full width at half-maximum (FWHM) $<$ 1 \AA] and weak (less than $5$
per cent below continuum) with central wavelengths that do not
correspond to known molecular or atomic species \citep{Herbig95,
  Hobbs09, Geballe11}. The originating physical sources (or
`carriers') of the DIBs have remained a source of speculation and
discussion since DIBs were first observed almost a century ago by
\citet{Heger22}.

\citet{Merrill34} originally suggested dust grains and molecules as
possible carriers for the DIB features. These carriers remain the most
plausible candidates after decades of observational, theoretical, and
experimental work (see reviews by \citealt{Herbig95},
\citealt{Fulara00}, and \citealt{Sarre06}). Substantial observational
evidence favors polyatomic, carbon-based molecular carriers. In
particular, photo-UV-resistant organic molecules \citep{Douglas77},
polycyclic aromatic hydrocarbons (PAHs) \citep{Cox11}, and fullerenes
\citep{Groth07} remain the most promising candidates. The recent
unambiguous identification of $C_{60}^+$ as the carrier for two
infrared DIB lines supports the notion that the carriers of DIBs are
indeed organic molecules
\citep{Campbell15}.

\begin{figure*}
\centering
\includegraphics[width=1\linewidth]{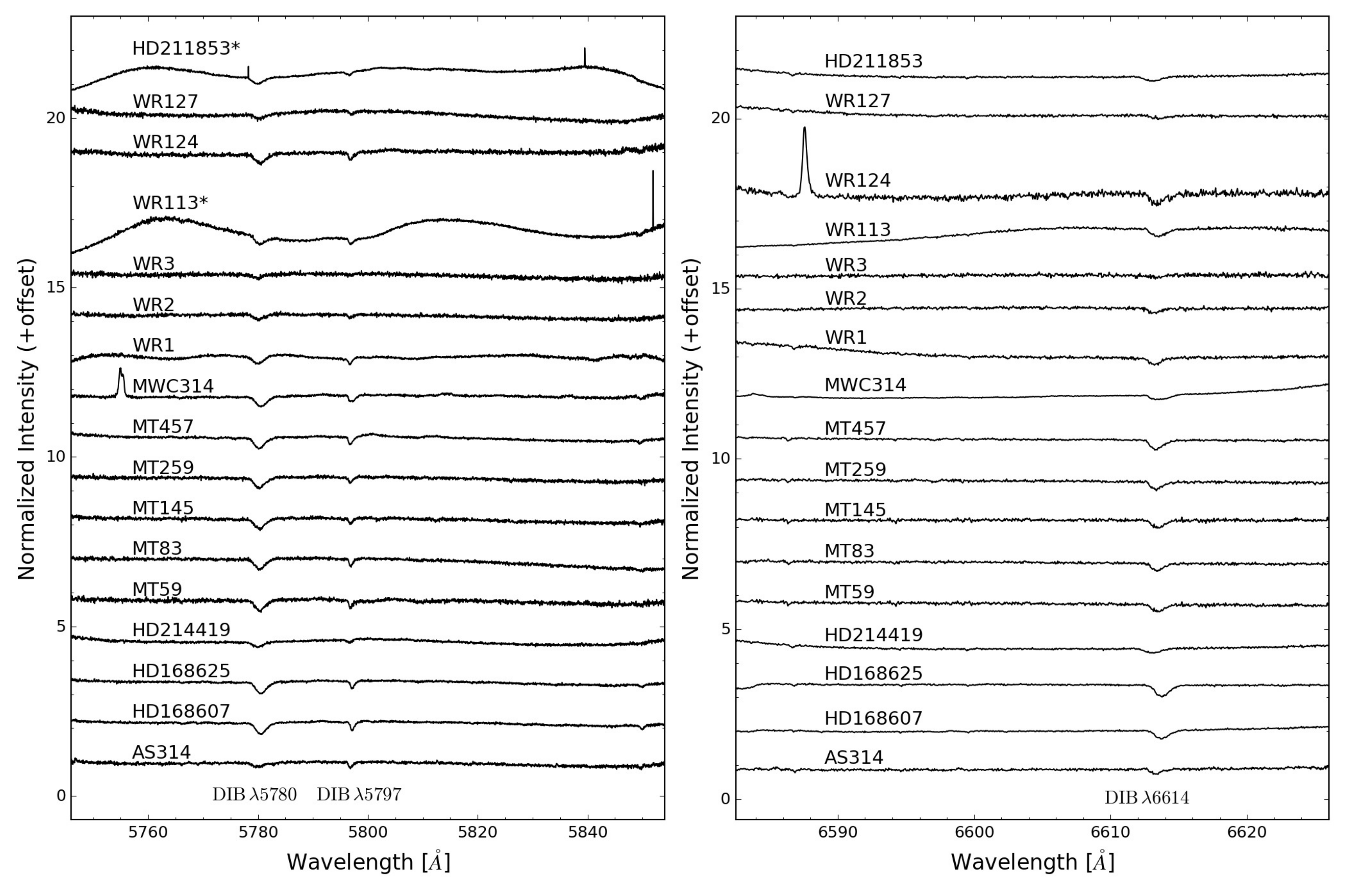}
\caption{Representative spectra of the 17 stars observed in our
  survey. Two echelle orders spanning DIB features of interest at
  5780, 5797, and 6614~\AA\ are shown. The complete data set
  are available online. Spectra of HD211853 and WR113 marked with an asterisk (`*') indicate instances where a 15th order Chebyshev function was insufficient to completely normalize the continuum.}

\label{fig:allspec}
\end{figure*}

Many observed properties of DIBs are best understood when
interpreted as originating in carriers located in the interstellar
medium (ISM). DIB features remain stationary in spectroscopic binaries
\citep{Merrill34}, and extinction and \ion {Na}{I} column density
positively correlate with DIB intensity \citep{Herbig95}.\\
\indent However, since the winds of massive stars are a principal
source of replenishment for the ISM, it has been suspected that DIB
carriers might also be present in circumstellar shells (CSs). To date,
despite some evidence for possible DIB-CS associations in a narrow
subset of mass-losing stars where temporal variations in DIB
equivalent width, central wavelength, and substructure were observed
\citep{Bertre90, Bertre93}, the majority of searches for carriers in
CSs have proven unsuccessful or inconclusive
\citep{Bertre92,Sarre06,Luna08}.

The recent discovery of time-varying DIB features in optical spectra
of the extragalactic Type Ic broad-lined supernova (SN) 2012ap calls
into question the current assumption that all DIBs reside in the ISM
\citep{Mili14}. The equivalent width (EW) strength of DIB features
observed toward SN\,2012ap changed with time in a manner consistent
with interaction between the SN and nearby DIB carrier(s) that possess
high ionization potentials like those found in small cations or
charged fullerenes.  \citet{Mili14} noted that archival data from two
additional spectroscopically similar SNe also show evidence for
temporal DIB variation, suggesting that a specific subset of SN
progenitor stars may possess a CS environment enriched with DIB
carriers.

Motivated by this recent discovery, we obtained high-resolution
multi-epoch optical spectra of 17 massive stars as part of a
pathfinder survey of variable DIBs. In this paper, we present
the results from our survey, which finds evidence for substructural
changes on short time-scales in the DIB absorption profiles towards
three stars. We describe the observations, reduction of the raw
spectral data, and the stellar sample selection in Section
\ref{sec:obs}. In Section \ref{sec:meth}, we discuss the Gaussian
fitting methods, determinations of errors, and data quality.  We use
these fitting methods and visual inspection to identify time-varying
changes in DIB absorption features in Section \ref{sec:results}. We
discuss how these findings may be interpreted as interaction between
strong stellar winds and the local carriers of DIBs in Section
\ref{sec:discussion} and summarize our conclusions in Section
\ref{sec:conclusions}.

\begin{figure}
\centering
\includegraphics[width=\linewidth]{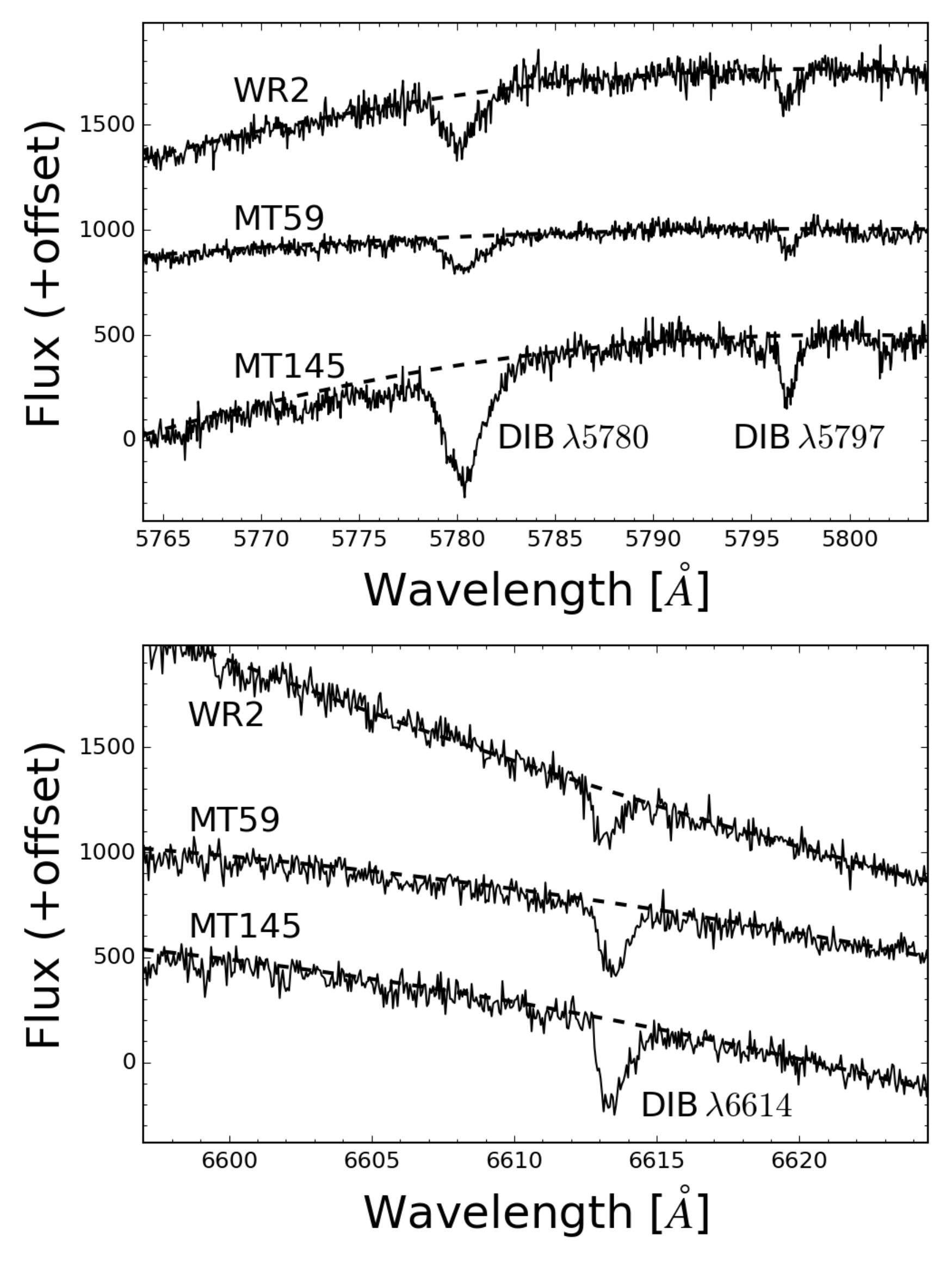}
\caption{Example fits to stellar continua of
  WR\,2, MT\,59, and MT\,145 used to normalize all spectra. For each
  epoch of these stars, the continuum was fitted with a 3rd order
  Chebyshev function to account for any continuum variability between
  observations. Fits to all other stars (see Table~\ref{tab:obs_list}
  for a complete listing) used also the Chebyshev function of order 3
  to 15.}

\label{fig:continuum}
\end{figure}

\begin{figure}
\centering
\includegraphics[width=\linewidth]{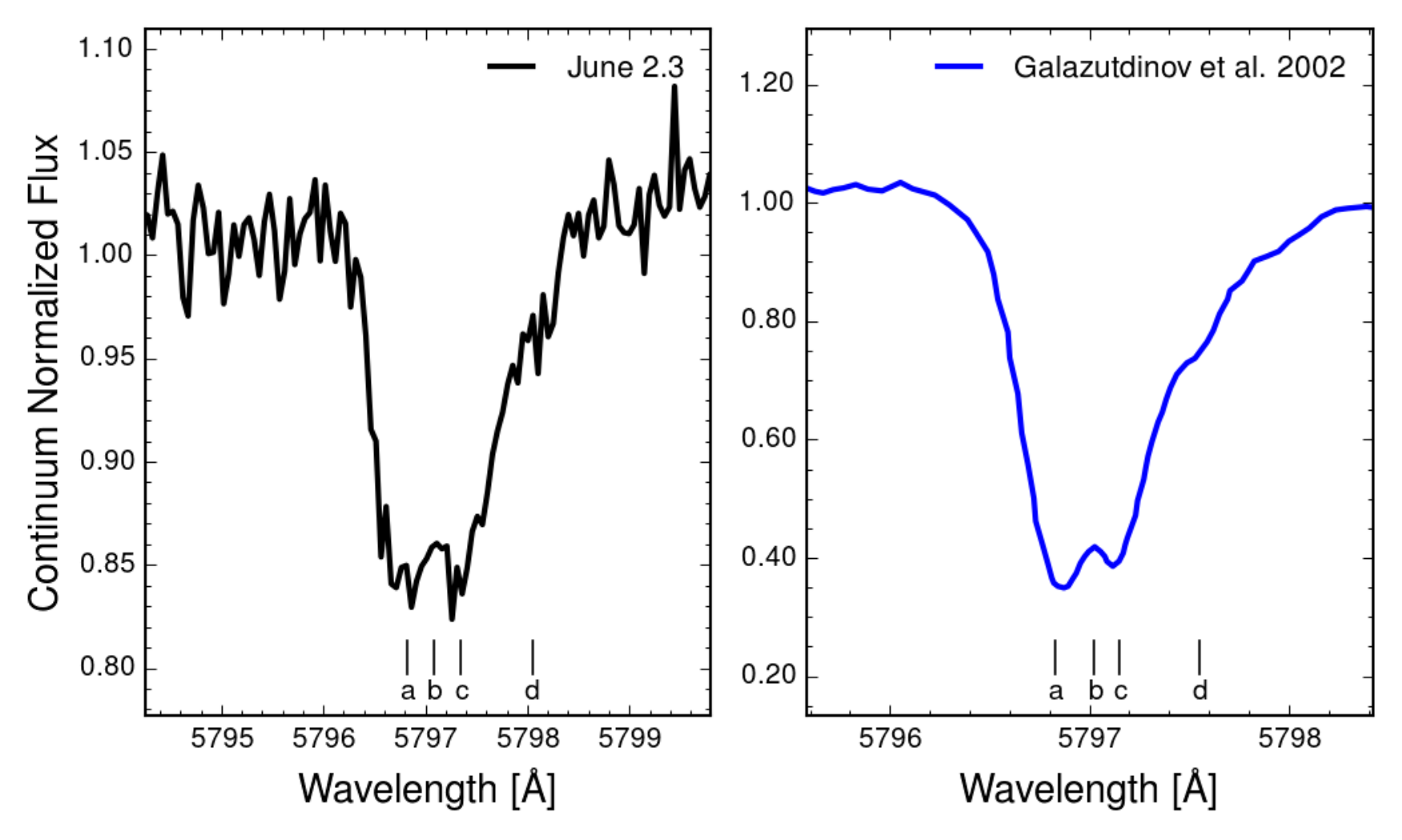}

\caption{Spectral comparison of DIB $\lambda5797$.
  The left plot shows a representative spectrum of MWC\,314 taken
  on June 2, 2013 that shows several substructural features in DIB
  $\lambda$5797. The right plot, a spectrum taken from \citet{Galazutdinov2002}
  using a spectrograph with resolution R$\sim$120,000, shows
  comparable substructure to the spectral data of MWC\,314.
  Four corresponding features have been labelled alphabetically in
  both plots.}

\label{fig:MWC314_sub_struct}
\end{figure}

\section{Observations}
\label{sec:obs}

\subsection{Instrument and Data}

High-resolution optical spectra were obtained with the 1.5-m
Tillinghast telescope and Tillinghast Reflector Echelle Spectrograph
(TRES) at Fred L. Whipple Observatory. TRES has a E2V 2K $\times$ 4.5K
CCD with 13.5-$\micron$ pixels, and each exposure provides nearly
complete spectral coverage from 3850 to 9100 \AA\, at a resolution of $R\approx 44\,000$ over the spectrograph's
51 orders \citep{Szent07, Mink11}. Multi-epoch observations were
obtained between June 2013 and January 2014 for each star with co-added total exposure times ranging from 300 to 1800 s depending on the
brightness of the source and observing conditions that were not always
photometric. Time between observations ranged from $\sim$10--60 d, were dependent on weather conditions and the
TRES observing queue, and thus were not obtained at regular
intervals. The typical spectral signal-to-noise (S/N) ratio in the
region of DIB features was $\ga$5--15.  A detailed list of
observations is found in Table \ref{tab:obs_list}.

\begin{table*}
\caption{Summary of observations.}
\label{tab:obs_list}
\resizebox*{!}{\textheight}{
\begin{tabular}{lcccccc} 
\hline \hline
Star & RA(2000) & Dec(2000) & Date & UTC & Exposure Time(s) & Air Mass \\ \hline
HD\,168607	& $18^{\circ} 21^{\prime} 15^{\prime \prime}$ & $-16^{\circ} 22^{\prime} 32^{\prime \prime}$	& 2013 June 2 &09:02:04& 360 & 1.49\\
&&& 2013 July 31	&05:33:41 & 450 & 1.50\\ \hline \\
HD\,168625	& $18^{\circ} 21^{\prime} 19^{\prime \prime}$ &$-16^{\circ} 22^{\prime} 26^{\prime \prime}$& 2013 June 2&09:10:20 & 360 & 1.50\\
&&& 2013 July 31&05:44:40 & 450 & 1.52\\ \hline \\
HD\,211853&$22^{\circ} 18^{\prime} 45^{\prime \prime}$ & $+56^{\circ} 07^{\prime} 33^{\prime \prime}$&2013 Sept 11&08:31:40 & 300 & 1.21\\
&&& 2013 Sept 19 & 07:22:19 & 300 & 1.15 \\
&&& 2013 Nov 14 & 04:42:23 & 300 & 1.25 \\
&&& 2013 Sept 25 & 04:39:53 & 300 & 1.11 \\ \hline \\
HD\,214419 & $22^{\circ} 36^{\prime} 53^{\prime \prime}$ & $+56^{\circ} 54^{\prime} 21^{\prime \prime}$ & 2013 Sept 11 & 08:59:27 & 300 & 1.23 \\
&&& 2013 Sept 19 & 07:30:01 & 300 & 1.15 \\
&&& 2013 Sept 23 & 03:30:26 & 360 & 1.21 \\
&&& 2013 Nov 14 & 04:49:42 & 300 & 1.23 \\ \hline \\
AS314 & $18^{\circ} 39^{\prime} 26^{\prime \prime}$ & $-13^{\circ} 50^{\prime} 47^{\prime \prime}$ & 2013 June 2 & 08:48:29 & 660 & 1.44 \\
&&& 2013 July 3 & 10:03:39 & 1000 & 2.07 \\
&&& 2013 July 31 & 06:04:58 & 900 & 1.45 \\ \hline \\
MT\,59 & $20^{\circ} 31^{\prime} 10^{\prime \prime}$ & $+41^{\circ} 31^{\prime} 53^{\prime \prime}$ & 2013 Sept 11 & 06:46:29 & 600 & 1.14 \\
&&& 2013 Sept 19 & 04:21:34 & 600 & 1.02 \\
&&& 2013 Sept 26 & 03:30:03 & 800 & 1.02 \\
&&& 2013 Nov 26 & 01:45:53 & 600 & 1.14 \\ \hline \\
MT\,83 & $20^{\circ} 31^{\prime} 22^{\prime \prime}$ & $+41^{\circ} 31^{\prime} 28^{\prime \prime}$ & 2013 Sept 11 & 07:01:14 & 1200 & 1.18 \\
&&& 2013 Sept 19 & 04:37:22 & 1200 & 1.03 \\
&&& 2013 Sept 25 & 03:19:07 & 1200 & 1.02 \\
&&& 2013 Nov 26 & 02:10:27 & 1200 & 1.20 \\ \hline \\
MT\,145 & $20^{\circ} 31^{\prime} 49^{\prime}$ & $+41^{\circ} 28^{\prime} 26^{\prime \prime}$ & 2013 Sept 19 & 06:00:58 & 1200 & 1.12 \\
&&& 2013 Nov 15 & 03:55:45 & 1200 & 1.42 \\ \hline \\
MT\,259 & $20^{\circ} 32^{\prime} 27^{\prime \prime}$ & $+41^{\circ} 28^{\prime} 52^{\prime \prime}$ & 2013 Sept 19 & 06:25:02 & 1200 & 1.17 \\
&&& 2013 Nov 15 & 04:18:45 & 1200 & 1.54 \\ \hline \\
MT\,457 & $20^{\circ} 33^{\prime} 14^{\prime \prime}$ & $+41^{\circ} 20^{\prime} 21^{\prime \prime}$ & 2013 Sept 19 & 06:48:15 & 1200 & 1.23 \\
&&& 2013 Nov 14 & 04:19:19 & 1200 & 1.52 \\ \hline \\
MWC\,314 & $19^{\circ} 21^{\prime} 33^{\prime}$ & $+14^{\circ} 52^{\prime} 56^{\prime}$ & 2013 June 02 & 08:18:15 & 1080 & 1.13 \\
&&& 2013 June 28 & 09:42:01 & 1200 & 1.12 \\
&&& 2013 July 24 & 08:21:08 & 900 & 1.15 \\ \hline \\
WR\,1 & $00^{\circ} 43^{\prime} 28^{\prime \prime}$ & $+64^{\circ} 45^{\prime} 35^{\prime \prime}$ & 2013 Sept 21 & 07:46:19 & 600 & 1.20 \\
&&& 2013 Sept 26 & 05:42:24 & 750 & 1.27 \\
&&& 2013 Nov 26 & 03:03:11 & 600 & 1.20 \\ \hline \\
WR\,2 & $01^{\circ} 05^{\prime} 23^{\prime \prime}$ & $+60^{\circ} 25^{\prime} 18^{\prime \prime}$ & 2013 Sept 11 & 10:05:30 & 1200 & 1.17 \\
&&& 2013 Sept 17 & 08:18:51 & 1200 & 1.14 \\
&&& 2013 Sept 21 & 08:00:49 & 1200 & 1.14 \\
&&& 2013 Sept 25 & 10:25:14 & 1350 & 1.25 \\
&&& 2013 Nov 26 & 03:23:14 & 1200 & 1.15 \\ \hline \\
WR\,3 & $01^{\circ} 38^{\prime} 56^{\prime \prime}$ & $+58^{\circ} 09^{\prime} 23^{\prime \prime}$ & 2014 Jan 11 & 03:07:16 & 600 & 1.16 \\
&&& 2014 Jan 21 & 02:48:53 & 600 & 1.19 \\ \hline
WR\,113 & $18^{\circ} 19^{\prime} 07^{\prime \prime}$ & $-11^{\circ} 37^{\prime} 59^{\prime \prime}$ & 2013 June 2 & 08:40:25 & 360 & 1.37 \\
&&& 2013 July 3 & 09:51:35 & 540 & 2.03 \\
&&& 2013 July 31 & 05:54:36 & 450 & 1.41 \\ \hline \\
WR\,124 & $19^{\circ} 11^{\prime} 30^{\prime \prime}$ & $+16^{\circ} 51^{\prime} 38^{\prime \prime}$ & 2013 June 2 & 07:52:29 & 1200 & 1.15 \\
&&& 2013 July 31 & 06:24:39 & 1200 & 1.04 \\ \hline \\
WR\,127 & $19^{\circ} 46^{\prime} 15^{\prime \prime}$ & $+28^{\circ} 16^{\prime} 19^{\prime \prime}$ & 2013 June 2 & 07:32:44 & 600 & 1.25 \\
&&& 2013 June 28 & 10:08:23 & 1800 & 1.07 \\
&&& 2013 July 27 & 07:52:31 & 1800 & 1.04 \\ \hline
\end{tabular}}
\end{table*}

Raw spectra were reduced using the standard TRES data reduction
pipeline \citep{Mink11}. The spectra were corrected for heliocentric
velocity using the \textsc{rvsao} software package in \textsc{iraf}
\citep{Kurtz98}. Continua of spectra were fit with Chebyshev functions
and then continuum normalized using the \texttt{continuum}
task. Example spectra of the 17 stars we observed around features of
interest is shown in Fig.~\ref{fig:allspec}. Fits to stellar
continua for three stars are shown in Fig.~\ref{fig:continuum}.
Orders used for fitting ranged from 3rd to 15th depending on the
complexity of the continuum. For each star, the lowest order fit over a broad range of wavelength in the relevant echelle order was chosen to mitigate any potential influence of other absorption features, which could skew the continuum fits. We have noted the few instances where the
continua were particularly complex and difficult to fit and we discuss
potential effects on our results in Section \ref{sec:discussion}. We also experimented with higher order fits but found no significant differences in the resulting continuum fits or candidate DIB changes between epochs.

\begin{figure*}
\centering
\includegraphics[width=\linewidth]{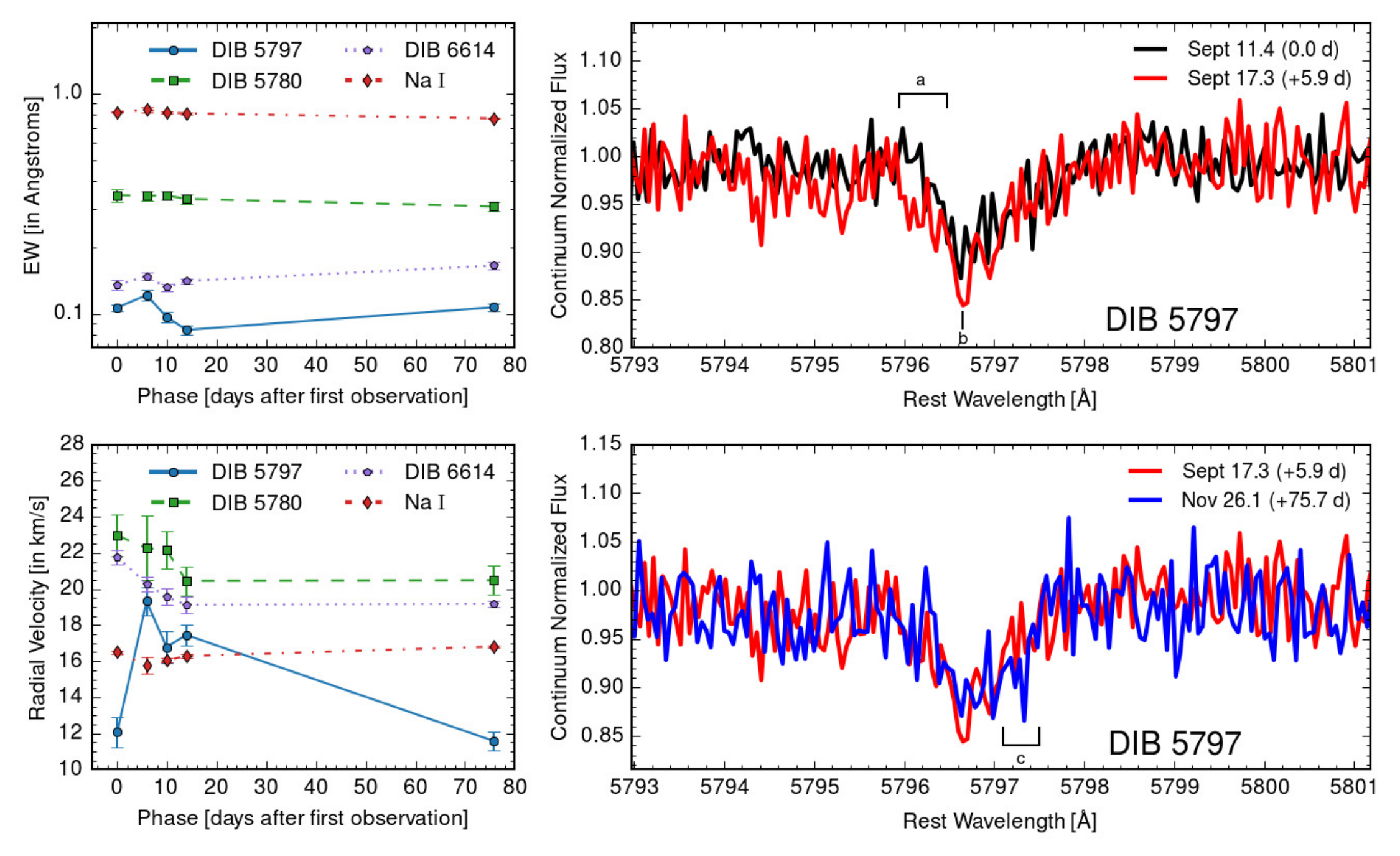}

\caption{Left: EW and radial velocity
  measurements for WR\,2. The features DIB $\lambda5797$, DIB
  $\lambda5780$, DIB $\lambda6614$, and \ion{Na}{I} are shown for all
  measured epochs. Note that the EW and radial velocity of \ion{Na}{I}
  remains stationary. Error bars represent $1\sigma$ uncertainties. Right: Multi-epoch comparison of raw spectra for DIB
  $\lambda5797$ feature of WR\,2. The plots step
  through selected epochs in which there is evidence for DIB
  substructural changes and the relevant wavelength regions are
  labelled alphabetically.}

\label{fig:WR2_changes}
\end{figure*}

\subsection{Sample Selection}

We observed 17 stars at multiple epochs for the DIB variability
survey. Candidate stars were selected from the literature based on
whether pre-existing evidence of DIBs existed, and if they were
mass-losing stars with enhanced O-rich or N-rich environments that
could be correlated with DIB variability (as suggested by
\citealt{Bertre93}). Wolf-Rayet (W--R) stars were favored since these
are believed to be the progenitor stars of Type Ib/c supernovae like
SN\,2012ap \citep{Gaskell86,Mili14}. Spectral type, extinction, and
further information for each source are found in Table
\ref{tab:star_info}.

\begin{table}
\caption{Properties of the stellar sample.}
\label{tab:star_info}
\resizebox{32em}{!}{\begin{tabular}{lcccccc} 
\hline \hline
Star & Spectral Class & Other Designations & $E(B-V)$ & Binarity & Reference \\ \hline
HD\,168607 & B9.4 Ia-0ep	&V4029 Sgr, MWC\,291	& 1.44& VB& 1, 2 \\
HD\,168625& B5.6 Ia-0p	&V4030 Sgr	&1.33& VB & 2\\
HD\,211853	& WN6o+06I 	& GP Cep	    &0.39&SB2& 3, 11, 13\\
HD\,214419& WN6+09II-Ib & CQ Cephei & 0.41&SB1& 4, 5, 11\\
AS\,314	& A0 Ia+ 	& V452 Scuti 	& 0.90&\ldots	& 6\\
MWC\,314 & B3Ibe	& V1429 Aquilae	& 1.45& SB1& 7  \\
WR\,1	& WN4-s	& HD\,4004	& 0.67&\ldots & 8, 14\\
WR\,2{*}	& WN2-w	& HD\,6327 	& 0.44&VB& 8, 14\\
WR\,113 & WC8d + 08-9IV	& HD\,168206, CV Ser & 0.78&SB2& 9, 11, 14\\
WR\,124 & WN8h& Hen 2-427	& 1.08&SB1& 8, 14\\
WR\,127	& WN3 + 09.5V	& HD\,186943 &0.27 &SB2& 8, 14\\
MT\,59{*}& O8.5 V & Schulte 1 &1.47 &SB1& 10, 12\\
MT\,83& B1 I & Schulte 2 & 1.18  &\ldots& 10, 12\\
MT\,145{*}& O9.5 V & Schulte 20 & 1.11 &SB1& 10, 12\\
MT\,259 & B0.5 V & Schulte 21 & 1.00 &\ldots& 10, 12\\
MT\,457 &O3 If & Schulte 7 & 1.45 &\ldots& 10, 12\\ \hline
\end{tabular}}
*Potential DIB variability detected. \\
\textbf{References} -- (1) \citealt{Chentsov04}; (2) \citealt{Walborn00}; (3) \citealt{Saurin10}; (4) \citealt{Skinner15}; (5) \citealt{Underhill90}; (6) \citealt{Miro00}; (7) \citealt{Carmona10}; (8) \citealt{Hamann06}; (9) \citealt{David-Uraz11}; (10) \citealt{Massey90}; (11) \citealt{Ducati02}; (12) \citealt{Kob14}; (13) \citealt{Panov90}; (14) \citealt{Hucht00}
\end{table}

\section{Methods and Analysis}
\label{sec:meth}

\subsection{Procedure}

Before individual exposures were co-added, spectra were
  visually examined and found to be free of noise spikes around the
  features of interest, which could produce spurious line profile
  variations. The DIB features at 5780.6, 5797.1, and 6613.7 \AA\
that were visible in all data were fit with Gaussian profiles using
\textsc{iraf}'s \texttt{splot} task. Gaussian profiles are not true
representations of DIB features but do provide a reasonable proxy by
which to estimate their strength and central wavelength. Due to the
poor response of the TRES instrument below 5000~\AA, the DIB
$\lambda$4428 feature is not detected in any of our observations. Some
stars weakly exhibited the 6196.0~\AA\ feature, but it could not be
reliably measured. The feature at 6283.9~\AA\ was also visible, but it
was often contaminated by telluric absorption and at the edge of the
order wavelength window. Thus, these features are not included in our
reported analysis. The \ion{Na}{I} feature, which provides a good
measure of wavelength stability between epochs and thus tests for
potential systematic errors, was fitted. We assumed 5889.950 and
5895.924~\AA\ for the intrinsic wavelengths \citep{MS73}. Multiple
fittings for the equivalent width (EW) were done for each feature and
averaged for each epoch, and the reported $1\sigma$
errors are given by the standard deviation of these fitting
results. Radial velocities were derived by comparing the central
wavelength of the Gaussian fittings to rest wavelengths of DIB
features taken from the NASA DIBs catalogue \citep{Jenniskens94}.

\subsection{Resolving DIB $\lambda$5797 Substructure}

We examined the well-studied DIB $\lambda 5797$ (cf.\
\citealt{Sarre95}) to assess our ability to measure possible variation
in DIB substructure. This DIB is known to have conspicuous
substructural features believed to be associated with the rotational
contours of gas-phase molecular carriers \citep{Kerr97}. We were able
to identify several characteristic substructural components in this
feature in all but two stars (AS\,314 and WR\,124). In these two outlying
cases, we were limited by poor S/N around the feature. A
representative spectrum showing the presence of this substructure is
shown in Fig. \ref{fig:MWC314_sub_struct}. This exercise
demonstrates that our observations with the TRES instrument permit
careful investigation of potential changes in DIB substructure.

\begin{figure*}
\centering
\includegraphics[width=\linewidth]{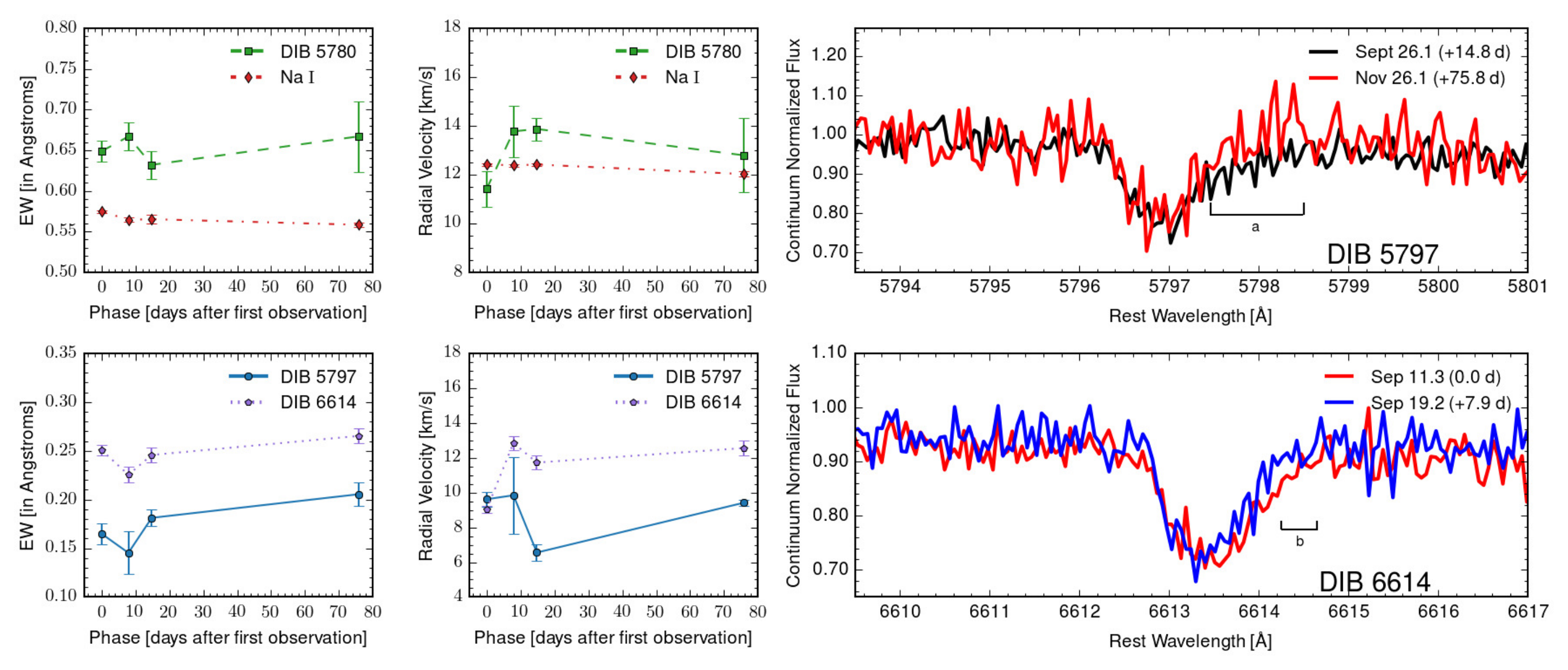}

\caption{Left: EW and radial velocity
  measurements for MT\,59. The features DIB $\lambda5780$, DIB
  $\lambda5797$, DIB $\lambda6614$, and \ion{Na}{I} are shown for all measured
  epochs. For clarity, DIB $\lambda5780$ is shown in the top row and
  DIB $\lambda5797$ is presented in the bottom row.  Note that the EW
  and radial velocity of \ion{Na}{I} remains stationary. Error bars represent $1\sigma$ uncertainties. Right:
  Multi-epoch comparison of raw spectra for DIB $\lambda5797$ and $\lambda6614$ features of
  MT\,59. The plots step through selected epochs in
  which there is evidence for DIB substructural changes and the
  relevant wavelength regions are labelled
  alphabetically. }

\label{fig:MT59_changes}
\end{figure*}

\section{Results}
\label{sec:results}

\subsection{Resolved DIB Features}

Our inspection of the data uncovered DIB variations in 3 of the 17
targets observed: WR\,2, MT\,59, and MT145. The remaining 14 stars
showed no measurable variations in DIB substructure, EW, or radial
velocity. Of the three candidate stars we identify DIB variability,
significant variations were seen in the 5797 and 6614 \AA\ features,
and marginal evidence for variation in the 5780 \AA\ feature. Below we
describe our results in detail.

\subsection{WR\,2}

WR\,2, also known as HD 6327, is a weak-lined Galactic WN2 star that
is compact ($0.89\,R_{\odot}$) and has a high temperature
($1.40 \times 10^5$ K). It exhibits rounded spectral profiles that are
unique among Galactic W--R stars and do not match current models
unless very rapid rotation ($\sim$1900 $\text{ km } \text{s}^{-1}$) is
assumed \citep{Hamann06}. For each epoch, the continuum was fitted
with a 3rd order Chebyshev function to account for any continuum
variability between observations. Examples of the continuum fits are
shown in Fig.~\ref{fig:continuum}. 

In Fig. \ref{fig:WR2_changes}, we plot the results of our
measurements, and show spectra of select epochs around the DIB
$\lambda5797$ feature illustrating how changes in radial velocity
coupled with modest EW variations are the consequence of substructural
changes within the DIB $\lambda5797$ feature. The radial velocity variations are seen at the $\approx 3\sigma$ level. The detection, disappearance, and
then slight re-emergence of a substructural feature on the blue wing
of the DIB as well as a deeper, sharply peaked feature in the blue side of the minimum of the absorption trough can be seen (labelled `a' and `b', respectively in Fig. \ref{fig:WR2_changes}). We also detect the appearance of a shallow dip in the right wing of DIB $\lambda5797$ (labelled `c' in Fig. \ref{fig:WR2_changes}) between September 17.3 and November 26.1. The DIBs at $5780$ and 6614 \AA\
show no measurable changes EW, and a marginal decrease in radial
velocity. \ion{Na}{I} remains unchanged throughout all epochs.

\begin{figure*}
\centering
\includegraphics[width=0.85\linewidth]{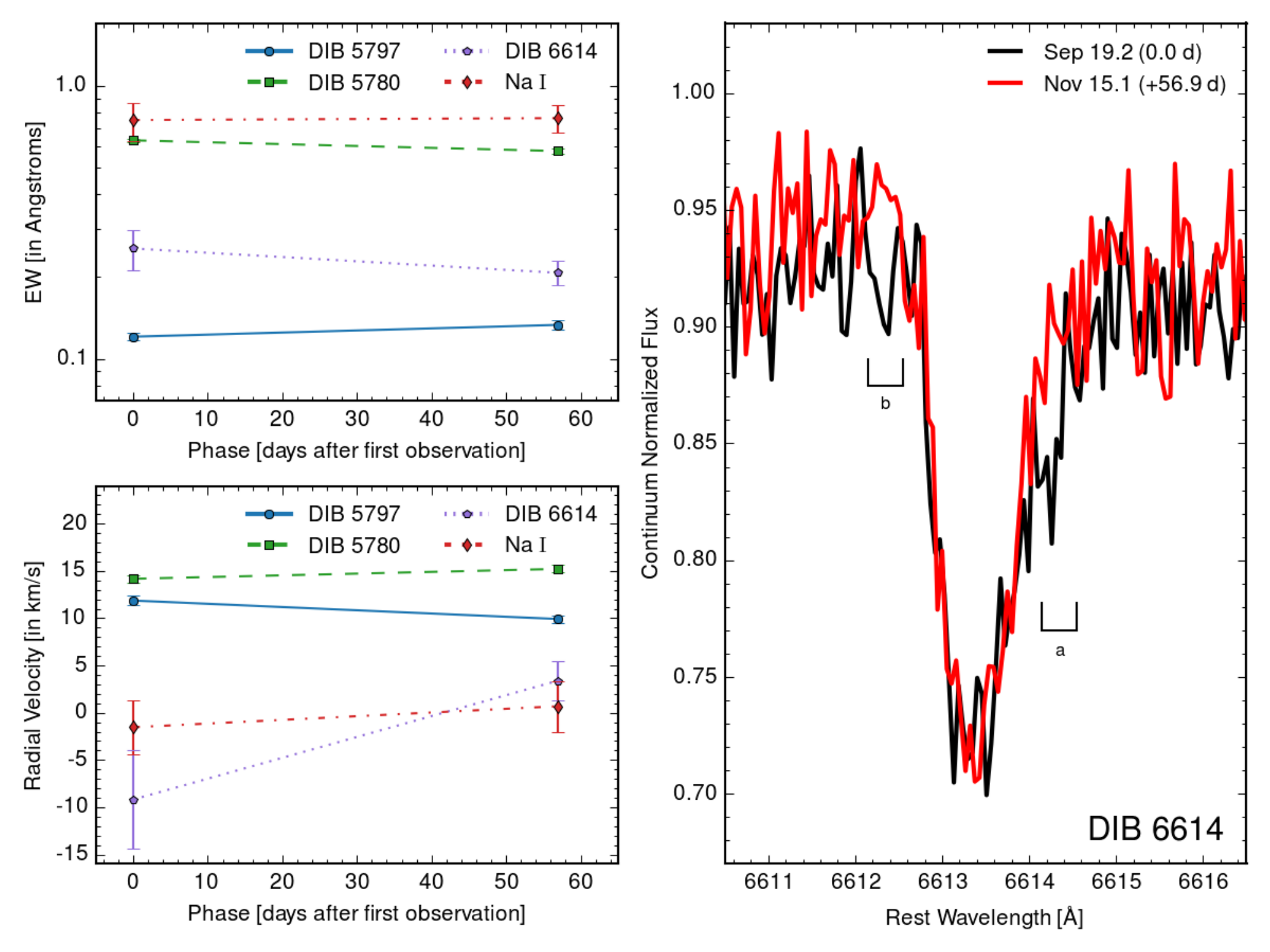}

\caption{Left: EW and radial velocity
  measurements for MT\,145. The features DIB $\lambda5780$, DIB
  $\lambda5797$, DIB $\lambda6614$, and \ion{Na}{I} are shown for all
  measured epochs. Note that
  the EW and radial velocity of \ion{Na}{I} remains stationary. Error bars represent $1\sigma$ uncertainties. Right:
  Multi-epoch comparison of raw spectra for DIB $\lambda 6614$ feature of
  MT\,145. The plots step through the two available
  epochs in which there is evidence for changes in DIB substructure. The
  one relevant wavelength region is labelled.}

\label{fig:MT145_changes}
\end{figure*}

\subsection{MT\,59}

MT\,59 is a massive O-type star located in Cygnus OB2 \citep{Massey90,
  Wright15}. For each epoch, the continuum was fitted with a 3rd order
Chebyshev function to account for any continuum variability between
observations (see Fig.~\ref{fig:continuum}).  In Fig.
\ref{fig:MT59_changes} we plot our measurements of changes in the DIB
$\lambda$5780, $\lambda$5797, and $\lambda$6614 features of MT\,59, as
well as select epochs of spectra enlarged around DIB $\lambda$5797 and
$\lambda6614$ features that exhibit the most conspicuous
changes. Between September 26.1 and November 26.1, a prominent substructural change is seen as well as a $\gtrsim3\sigma$ variation in the radial velocity of DIB $\lambda$6614. The red wing of the DIB spanning 5797.6 to 5798.6 \AA\ (labelled
`a' in Fig.~\ref{fig:MT59_changes} and labelled `d' in
Fig.~\ref{fig:MWC314_sub_struct}) decreases in strength. A marginal decrease is also detected in the red wing of the DIB $\lambda$6614 feature (labelled `b' in Fig. \ref{fig:MT59_changes}) between September 11.3 and September 19.2. The DIB at 5780 \AA\ shows no measurable changes in EW and negligible variations in radial velocity. \ion{Na}{I} remains unchanged throughout all epochs.

\subsection{MT\,145}

MT\,145 is also a massive O-type star located in Cygnus OB2
\citep{Massey90, Wright15}. For each epoch, the
continuum was fitted with a 3rd order Chebyshev function to account
for any continuum variability between observations.  Measurements of
changes in its DIB $\lambda$5780, $\lambda$5797, and $\lambda$6614
features are shown in Fig. \ref{fig:MT145_changes}, as well as
spectra enlarged around DIB $\lambda$6614 that exhibited the most
conspicuous change. From September 19.2 to November 15.1, we detect an increase in a broad feature on the red and left wings of DIB $\lambda$6614 (labelled `a' and `b' in Fig. \ref{fig:MT145_changes}, respectively). The DIBs at 5780 and 5797 \AA\ exhibit no measurable changes in EW and only minor radial velocity changes. \ion{Na}{I} remains unchanged throughout all epochs.

\subsection{Robustness of Our Detections}

In order to quantify the significance of our DIB temporal
  variation detections, we compared the average flux excess (or
  deficiency) as measured by the difference between our spectra and
  continuum fits. The difference between observation and continuum fit
  was made in two directly adjacent 2.5 \AA\ windows centered on
  the DIB feature. An advantage of this method is that it incorporates
  the relatively broad spectral width of the observed variations that
  are much larger than narrow pixel-to-pixel noise fluctuations. For
  instance, the substructural feature labeled `a' in
  Fig. \ref{fig:WR2_changes} is not substantially larger in flux than
  a noise feature at about 5794.3 \AA\ but `a' is distributed over
  $\approx0.6$ \AA\ while the noise feature of similar depth happens
  only over $\sim0.2$ \AA\ ~.

For profiles consistent with no temporal change, no
  substantial flux excess is expected relative to that of the
  continuum, which is set by the root-mean-square
  fluctuations. However, for the broad changes seen toward the wings
  of DIBs profiles, which represent the majority of our candidate
  changes, we typically observe $\sim1.5\times$ flux differential between the DIB
  features and the continuum. The variation observed in WR\,2 in DIB
  $\lambda$5797 between Sept. 11.4 and 17.3 has a particularly large
  flux excess that is factor of $\sim24\times$ larger than the
  continuum. We note that this method is not sensitive to changes in
  the form of narrow spectral spikes, e.g., WR 2, `c' for DIB
  $\lambda$5797, that may in fact be attributable to noise in the
  extracted spectrum.

Variable DIB features detected in WR\,2 and MT\,59 pass our
  flux excess test and support the view that the observed changes are
  intrinsic to the profile and not artificial. The candidate detections in
  MT\,145, however, do not pass this test and thus we are less
  confident in their reality.

\section{Discussion}
\label{sec:discussion}

\subsection{Variable Diffuse Interstellar Bands}

We observe measurable changes in absorption line substructure around
the DIB $\lambda$5797 and $\lambda6614$ features over relatively short
time-scales ($t < 60$ d). These changes are unlike normal DIB profiles
that are static and remain unchanged, even after $> 40$ yr of
observing \citep{Herbig95}. They are also unlike the changes observed
in DIB absorptions toward SN\,2012ap \citep{Mili14}. In that case, the
evolution was observed as pronounced changes in EW absorption
strengths, as opposed to modifications of profile substructure that we
observe in WR\,2, MT\,59, and possibly MT\,145.

Robust fitting of stellar continua is a known challenge in DIB profile
analyses (see, e.g., \citealt{Galazutdinov2008}), and could
potentially contribute to changes in DIB profile structure like those
seen in our spectra. We are unable to conclusively rule out a
  changing continuum between observational epochs. However, no
conspicuous changes in the continua that have been fit with 3rd order
Chebyshev polynomials between epochs is observed. Thus, we interpret
these changes to be intrinsic to the DIB profiles and not due to
changes in the stellar continua.

Spectra of increasing quality and resolution obtained over the past
two decades have hinted at variations in the profiles of individual
DIBs along different sight lines. In some cases, the observed profiles
reflect the complex structure of the ISM made up of multiple clouds
moving at a range of velocities (e.g., \citealt{Herbig82};
\citealt{Weselak10}). However, not all differences in profile shapes
observed along different sight lines can be accounted for this
way. Moreover, defining characteristics of these profile shape
differences resemble the DIB variations reported here.

For example, \citet{Krelowski1997} observed the DIB $\lambda5797$
feature at $R\approx60\,000$ using the McDonald Observatory echelle
spectrograph toward several $\zeta$, $\rho$, and $\sigma$-type sources
(see their paper for a more careful definition of these terms), and
observed slight differences in the fine structure on the redward side.
These subtle substructure variations have been repeatedly documented
in a series of high-resolution surveys of sight lines probing only a
single cloud by \citet{Galazutdinov2002}, \citet{Galazutdinov2008} and
\citet{Galazutdinov08MNRAS}, which confirmed that the variations in
profile are intrinsic to the carrier, and not due to Doppler effects.

\begin{figure}
\centering
\includegraphics[width=0.8\linewidth]{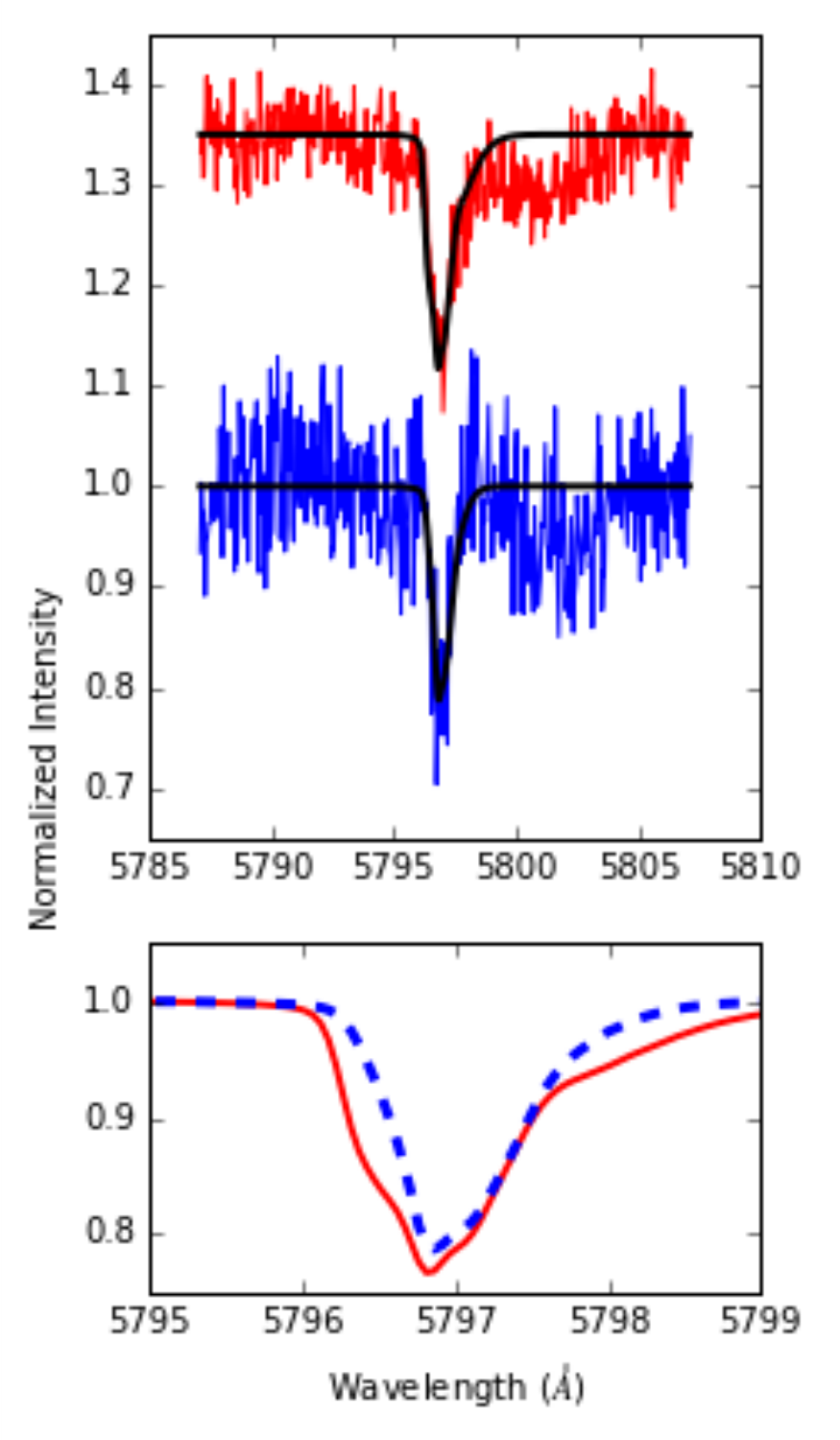}

\caption{Model fits to the DIB $\lambda$5797 profile of MT 59 using the models of \citet{Huang2015}. The top panel shows the observational data and best model for September 26 (top, Model A, 60\% 45~K) and November 26 (bottom, Model A, 50\% 20~K). The bottom panel contains a closer view of the differences between the models. }

\label{fig:model_fits}
\end{figure}

\citet{Dahlstrom2013} and \citet{Oka2013} have reported observations
of several DIBs ($\lambda$5780, $\lambda$5797, and $\lambda$6614) that
show very prominent extended tails toward the red (ETRs) in Her 36,
suggesting that the carriers are polar molecules that are being
radiatively pumped by nearby dust emission, leading to elevated
rotational temperatures.  A comparison of their spectra toward Her 36
and 9 Sgr (which shows no ETR), shown in fig.~3 of \citet{Oka2013},
bears striking resemblance to the comparison of our spectra toward MT
59 taken on September 26 and November 26, respectively
(Fig.~\ref{fig:MT59_changes}). Likewise, the differences observed in
the extended sidebands of the DIB $\lambda 6614$ profiles towards HD
179406 and HD 147889 presented in \citet{Marshall15} and explained in
terms of vibrational hot bands, are similar to the two epochs of
MT\,145 presented here (Fig.~\ref{fig:MT145_changes}). 

Searches for time-varying DIBs were conducted toward
  $\kappa$~Velorum by \citet{Smith13}. Known atomic absorption lines
  changed over time but DIB features did not, indicating that DIBs do
  not necessarily follow the same trends as atomic transitions. These
  results are not inconsistent with our own findings. Whereas the DIBs
  investigated in \citet{Smith13} were primarily interstellar in
  origin, we propose that the variations we observe must result from
  circumstellar DIB carriers interacting with the massive star winds.

These results all support the notion that DIB absorption profiles can
exhibit substructure that is sensitive to the environmental conditions in which
carrier molecules reside. In the next section we test the
plausibility of this hypothesis by modeling the variable DIB $\lambda
5797$ profile of MT\,59 assuming changes in temperature.

\subsection{DIB $\lambda$5797}

DIB $\lambda$5797 is a strong, well-studied feature whose profile
resembles the rotational fine structure associated with a molecular
rovibronic transition, as best illustrated by the $R\approx600\,000$
spectrum of \citet{Sarre95} acquired toward $\mu$ Sgr using the
Ultra-High-Resolution Facility at the Anglo-Australian
  Telescope \citep{Diego95}.  Our observations of MT\,59 may suggest
that at least some of the molecular carriers of DIB $\lambda$5797 are
experiencing time-varying internal excitation due to an interaction
with the star.  If this is true, the nature of the interaction may
help shed light on the molecular carrier. However, since the carrier
is unknown, it is difficult to reliably interpret the changes in the
profile.

Several attempts have been made to model
the fine structure of DIBs $\lambda$5797 in terms of specific types of
spectroscopic transitions of free molecules.
\citet{Kerr97} performed a contour modeling study based on a
perpendicular transition in a planar (oblate) symmetric top molecule,
but were unable to achieve a satisfactory match.   \citet{Oka2013} 
modeled the ETRs as a
$^2\Pi \leftarrow ^2\Sigma$ perpendicular band in a linear molecule.
Although their model reproduced the ETRs, it did not match the fine
structure at the top of the band.  A later model of DIB $\lambda$5797
in a more typical source by \citet{Huang2015}, based on a
$^2\Pi \leftarrow ^2\Pi$ parallel transition in a linear radical with
5--7 heavy atoms, was able to generally reproduce the ETRs with some
success assuming a 2 component mixture with different rotational
temperatures in order to approximate a higher level of rotational excitation due to radiative pumping. The same model can also generally account for the
structure in more typical sources that do not display the ETRs by
assuming a rotational temperature of 2.73 K appropriate for a polar
molecule in a low-density environment.

The limited resolution of our spectra and low S/N
$\approx 5$--$15$ of DIB $\lambda$5797 in our data preclude a detailed
quantitative analysis of the profile variations.  However, if we adopt
the interpretation of \citet{Huang2015}, we might ascribe the increase
in absorption on the red side of this feature in MT\,59 (Fig.
\ref{fig:MT59_changes}) to a time-dependent change in the excitation
temperature of a portion of the DIB carriers due to an interaction
with the massive star.  Such an effect could occur in one of two ways:
1) if the radiation field of the star changes with time, a portion of
the carriers could be radiatively pumped into higher internal states,
ultimately leading to a higher rotational excitation temperature, or
2) some of the DIB carriers may experience an interaction with a
clumpy mass-loss wind, leading to an increased collision rate that
increases the excitation temperature of the carrier's rotational
distribution above 2.73 K.  

We inspected the stellar features of our candidate stars for
conspicuous and unexplainable changes coincident with changes observed
in the DIB features but were unable to find any. Specifically, the
fine structure of the emission line profiles of WR\,2 did not exhibit
any measurable changes between epochs, and the temporal variation of H
and He absorption features of MT\,59 and MT\,145 (most conspicuous
being H\,$\alpha$ and \ion{He}{I} $\lambda$5876) was consistent with
their nature as single-lined binary systems \citep{Kob14}. Thus we
have no direct evidence that the stellar radiation field exhibited
significant changes during the observational period. However, massive
stars are known to exhibit temporally varying mass-loss winds on these
time-scales \citep{Moffat88,Tuthill99}, so the latter interpretation
is attractive.

We attempted to reproduce the DIB $\lambda$5797 profiles of MT\,59 using the program PGOPHER\footnote{http://pgopher.chm.bris.ac.uk/}, in combination with the models of \citet{Huang2015}. For each model (A, B, and C), molecular spectra were simulated at temperatures ranging from the background radiative temperature (2.73~K) up to 60~K. We then constructed synthetic spectra from two-component mixtures consisting of a 2.73~K component with variable amounts (0--100\%) of a second, higher temperature component for each original model from~\citet{Huang2015} in order to simulate additional rotational excitation. Comparing all these models to the MT\,59 September 26 data, the best match involved a $\sim$50/50 mixture of a 40--60~K and a 2.73~K spectrum derived from Model A, while the MT\,59 November 26 was best fit by a 50/50 mixture whose warmer component was 20 K (Figure \ref{fig:model_fits}). Model C showed qualitatively similar results, while model B was found to be a poor fit in general to the September 26 data. While higher SNR spectra will be required to derive any firm conclusions, our preliminary analysis supports the notion that a change in the rotational excitation of the DIB carrier population could explain the profile changes tentatively observed between epochs.

\section{Conclusions}
\label{sec:conclusions}

We have presented the results of a pathfinder survey of 17 massive
stars conducted with the TRES high resolution spectrograph that looked
for changes in DIB absorption features over multiple epochs. We find
evidence for time-varying changes in three stars WR\,2, MT\,59, and
MT\,145 in the substructure of the DIB $\lambda$5797 and $\lambda$6614
features over relatively short time-scales ($t < 60$ d).  The changes
are most pronounced along the wings of the profiles. Sharp and narrow
features ($<0.2$ \AA) are also seen to come and go, but their reality
is less secure. The changes in WR\,2 and MT\,59 pass our flux excess
test and are our leading candidate detections, whereas the changes in
MT\,145 fail this test and have a greater chance of being due to noise
in the extracted spectrum.

We have proposed that the short-term DIB variability we observe may be
the consequence of carrier molecules interacting with the winds of
nearby host stars. We do not find any correlation between DIB changes
and changes in stellar emission/absorption features to support this
hypothesis. However, using the models of \citet{Huang2015}, we find it
plausible that changes in rotational excitation of the carrier molecules
could be responsible for the observed changes in DIB substructure
observed in our data between epochs.

Our results imply that carrier molecules responsible for the DIB
absorption variability must be located in close proximity to massive
stars (i.e., within their circumstellar shells). This conclusion runs
counter to numerous studies that support the view that DIBs are a
static large-scale phenomenon not located in circumstellar
environments \citep{Seab95,Luna08,Cox11}. However, theoretical and
observational evidence increasingly favor the notion that DIB carriers
are large gas phase molecules (e.g. PAHs or fullerenes) that could
very well be present in circumstellar shells and detectable under
specific circumstances.

Notably, recent work suggests that the mere presence of carbon-rich
space environments may not be enough to detect DIB carrier material,
and that proximity to a strong source of ionizing photons may be an
important factor in producing enhanced DIB absorptions (see, e.g.,
\citealt{Mili14}; and \citealt{DiazLuis15} who report evidence of
unusually intense DIB $\lambda$4428 and DIB $\lambda$5780 features in
fullerene planetary nebulae). Furthermore, if mass loss is
concentrated along circumstellar discs, then the ability to observe DIB
variations may also be related to line of sight effects. Taken
together, observing DIBs in circumstellar environments may be
relatively difficult and the instances rare given the numerous
conditions that need to be satisfied in order for them to be
detectable.

Our survey demonstrates the need for instrumental stability,
appropriate observing cadence, high spectral resolution, and superior
S/N in order to successfully detect DIB profile changes.  A
future survey could test our claims of DIB variability and improve on our work by observing more stars
with spectra of S/N $\gtrsim$ 100. Observations made at evenly spaced intervals matching orbital periods of known binary systems could look for repeated DIB variability correlated with stellar absorption and/or emission
features. Broader wavelength coverage could 
potentially identify new families of time-varying DIBs and probe the
DIB $\lambda 4428$ feature, which was not studied in this
investigation but exhibited the most conspicuous changes in spectra
observed from SN\,2012ap. Using
information about the environments of these stars (e.g.\ temperature
and ionizing radiation) in combination with contour models, the
temporal variations observed in these DIBs may provide a new way to
probe the physical properties of their carriers.

\section*{Acknowledgements}

We thank I.\ Crawford who provided many helpful comments and
  suggestions that improved the quality and presentation of this
  paper. E. Berger kindly provided helpful comments on an earlier
draft of the paper. The observations reported in this paper were
obtained at the Fred L.\ Whipple Observatory, which is operated by the
Smithsonian Astrophysical Observatory. We thank L.\ Buchhave, G.\
Esquerdo, P.\ Berlind, M.\ Calkins, and J.\ Mink for help planning,
obtaining, and reducing these data. C.~L.\ thanks the Harvard College
Program for Research in Science and Engineering that provided
financial support for this work, as well as the Universities Space
Research Association for additional support through the Frederick
Tarantino Memorial Scholarship Award.




\bibliographystyle{mnras}
\bibliography{mybib2} 



\bsp	
\label{lastpage}
\end{document}